\documentclass[twocolumn,trackchanges]{aastex61}

\received{March 20, 2017}
\accepted{June 8, 2017}

\shorttitle{}
\shortauthors{Abell\'an et al.}

\begin{document}

\title{Very deep inside the SN 1987A core ejecta: molecular structures seen in 3D}

\correspondingauthor{F. J. Abell\'an}
\email{francisco.abellan@uv.es}

\author[0000-0002-0786-7307]{F. J. Abell\'an}
\affil{Departamento de Astronom\'ia y Astrof\'isica, Universidad de Valencia, C/Dr. Moliner 50, ES-46100 Burjassot, Spain}

\author{R. Indebetouw}
\affiliation{Department of Astronomy, University of Virginia, PO Box 400325, Charlottesville, VA 22904, USA}
\affiliation{National Radio Astronomy Observatory, 520 Edgemont Rd, Charlottesville, VA 22903, USA}

\author{J. M. Marcaide}
\affil{Departamento de Astronom\'ia y Astrof\'isica, Universidad de Valencia, C/Dr. Moliner 50, ES-46100 Burjassot, Spain}

\author{M. Gabler}
\affiliation{Max-Planck-Institut f\"ur Astrophysik, Karl-Schwarzschild-Stra\ss e 1, D-85748 Garching, Germany}

\author{C. Fransson}
\affiliation{Department of Astronomy, The Oskar Klein Centre, Stockholm University, Alba Nova University Centre, SE-106 91 Stockholm, Sweden}

\author{J. Spyromilio}
\affiliation{ESO, Karl-Schwarzschild-Stra\ss e 2, D-85748 Garching, Germany}

\author{D. N. Burrows}
\affiliation{Department of Astronomy $\&$ Astrophysics, The Pennsylvania State University, University Park, PA 16802, USA}

\author{R. Chevalier}
\affiliation{Department of Astronomy, University of Virginia, PO Box 400325, Charlottesville, VA 22904, USA}

\author{P. Cigan}
\affiliation{School of Physics and Astronomy, Cardiff University, Queen’s Buildings, The Parade, Cardiff, CF24 3AA, UK}

\author{B. M. Gaensler}
\affiliation{Dunlap Institute for Astronomy and Astrophysics, University of Toronto, Toronto ON M5S 3H4, Canada}

\author{H. L. Gomez}
\affiliation{School of Physics and Astronomy, Cardiff University, Queen’s Buildings, The Parade, Cardiff, CF24 3AA, UK}

\author{H.-Th. Janka}
\affiliation{Max-Planck-Institut f\"ur Astrophysik, Karl-Schwarzschild-Stra\ss e 1, D-85748 Garching, Germany}

\author{R. Kirshner}
\affiliation{Harvard-Smithsonian Center for Astrophysics, 60 Garden St., Cambridge, MA 02138, USA}

\author{J. Larsson}
\affiliation{KTH, Department of Physics, and the Oskar Klein Centre, AlbaNova, SE-106 91 Stockholm, Sweden}

\author{P. Lundqvist}
\affiliation{Department of Astronomy, The Oskar Klein Centre, Stockholm University, Alba Nova University Centre, SE-106 91 Stockholm, Sweden}

\author{M. Matsuura}
\affiliation{School of Physics and Astronomy, Cardiff University, Queen’s Buildings, The Parade, Cardiff, CF24 3AA, UK}

\author{R. McCray}
\affiliation{Department of Astronomy, University of California, Berkeley, CA 94720-3411, USA}

\author{C.-Y. Ng}
\affiliation{Department of Physics, The University of Hong Kong, Pokfulam Road, Hong Kong, China}

\author{S. Park}
\affiliation{Department of Physics, University of Texas at Arlington, 108 Science Hall, Box 19059, Arlington, TX 76019, USA}

\author{P. Roche}
\affiliation{Department of Physics, University of Oxford, Oxford OX1 3RH, UK}

\author{L. Staveley-Smith}
\affiliation{Australian Research Council, Centre of Excellence for All-sky Astrophysics (CAASTRO)}
\affiliation{International Centre for Radio Astronomy Research (ICRAR), University of Western Australia, Crawley, WA 6009, Australia}

\author{J. Th. Van Loon}
\affiliation{Lennard-Jones Laboratories, Keele University, ST5 5BG, UK}

\author{J. C. Wheeler}
\affiliation{Department of Astronomy, University of Texas, Austin, TX 78712-0259, USA}

\author{S. E. Woosley}
\affiliation{Department of Astronomy and Astrophysics, University of California, Santa Cruz, CA 95064, USA}



\begin{abstract}

Most massive stars end their lives in core-collapse supernova explosions and enrich the interstellar medium with explosively nucleosynthesized elements. Following core collapse, the explosion is subject to instabilities as the shock propagates outwards through the progenitor star. Observations of the composition and structure of the innermost regions of a core-collapse supernova provide a direct probe of the instabilities and nucleosynthetic products. SN 1987A in the Large Magellanic Cloud (LMC) is one of very few supernovae for which the inner ejecta can be spatially resolved but are not yet strongly affected by interaction with the surroundings. Our observations of SN 1987A with the Atacama Large Millimeter/submillimeter Array (ALMA) are of the highest resolution to date and reveal the detailed morphology of cold molecular gas in the innermost regions of the remnant. The 3D distributions of carbon and silicon monoxide (CO and SiO) emission differ, but both have a central deficit, or torus-like distribution, possibly a result of radioactive heating during the first weeks (``nickel heating''). The size scales of the clumpy distribution are compared quantitatively to models, demonstrating how progenitor and explosion physics can be constrained.
\end{abstract}




\section{Introduction} \label{sec:intro}

Supernova 1987A in the LMC has provided an excellent laboratory for supernova physics. Its short distance (50 kpc) has provided the very rare opportunity to spatially resolve the supernova as it evolves (currently the outer shock is greater than $1\farcs5$ or $\sim1$ light-year in diameter). The environment of the ejecta of SN 1987A consists of a thin dense circumstellar ring, the \textit{equatorial ring}, that is tilted 43$^{\circ}$ from the line of sight \citep{Jakobsen91, Tziamtzis11}. The first evidence of interaction of the outermost ejecta with the equatorial ring appeared in 1995 \citep{Sonneborn98, Lawrence00}, and was followed by the emergence of hotspots along the equatorial ring in the following years. These shocks and the reverse shock have provided much of the energy for the recent electromagnetic display from the supernova, as the decay of the radioactive energy sources produced in explosive nucleosynthesis provides ever decreasing amounts of energy. 

At present, the ejecta emission at optical wavelengths is dominated by gas illuminated by the X-ray radiation from the ring \citep{Larsson11, Fransson13}. Our view of the inner debris where most of the nucleosynthesis products reside is highly obscured at optical/near-infrared by a massive cloud of dust grains \citep{Matsuura15, McCray16}. By contrast, millimeter molecular emission originates in the cold innermost ejecta and is free of the obscuration, thus being particularly powerful for understanding the details of elemental distribution and mixing. 

Elements in the ejecta material are not simply radially stratified as might be expected from stellar evolution. Theory suggests that instabilities mixed the stellar nucleosynthetic products into a clumpy 3D structure. These began during the neutrino-driven phase of the explosion by convective overturn \citep{Burrows95} and/or standing accretion shock instability \citep[SASI,][]{Blondin03}, and grew by Rayleigh-Taylor instabilities as the shock propagated outwards through the progenitor during the first $\sim10^4$ s \citep{Ebisuzaki89, Benz90}. During the next days, radioactive decay of $^{56}$Ni$\rightarrow^{56}$Co$\rightarrow^{56}$Fe heats clumps rich in those heavy elements, causing some additional expansion of those clumps. If the process is sufficiently important, a large-scale reduction of heavy products from the central regions and lowest velocities may occur \citep[``nickel heating'',][]{Woosley88, Herant91, Herant92, Basko94}. After the first few days, this structure was frozen in homologous free expansion, a record that can be analyzed long afterwards. 

Early detection of hard X-rays and the smooth shape of the light curve provided observational evidence that freshly synthesized material from the core of the supernova had found its way to the outer, less optically thick, regions of the envelope \citep{Arnett89, McCray93}. Observations of the structure of atomic emission lines \citep{Haas90, Spyromilio90} also suggested that such macroscopic mixing of the ejecta had taken place. Detailed spectral modeling \citep[and references therein]{McCray93}, however, placed limits on the microscopic elemental mixing. Adiabatic expansion and radiative cooling caused temperatures to drop rapidly and allowed molecules to form where different atoms coexist. The molecular excitation temperature of $20-170$ K \citep{Kamenetzky13, Matsuura17} reflects the balance between this cooling and heating due to the gamma rays and leaking positrons from the $^{44}$Ti decay, and possible X-rays and UV emission from the ring. Vibrational emission of CO and SiO was observed within the first 2 years \citep{Rank88,Spyromilio88,Roche91}, and provided further constraints on the amount of mixing. However, none of these observations spatially resolved the heterogeneous ejecta. Modest angular resolution ($\sim500$ mas) ALMA data \citep{Kamenetzky13} revealed the presence of bright rotational emission from cold ($<150$ K) relatively slow moving ($\sim2000$ km s$^{-1}$) carbon monoxide CO and silicon monoxide SiO. 

We have now observed SN 1987A with ALMA in the CO $J=2-1$, SiO $J=5-4$ and $J=6-5$ rotational transition lines using the long baseline, high angular resolution mode of the telescope. These observations are of the highest resolution ever taken for SN 1987A and enable us to create and compare the first 3D maps of molecular emission from deep inside of the remnant with state-of-the-art explosion models. 

\begin{figure*}[ht!]
\plotone{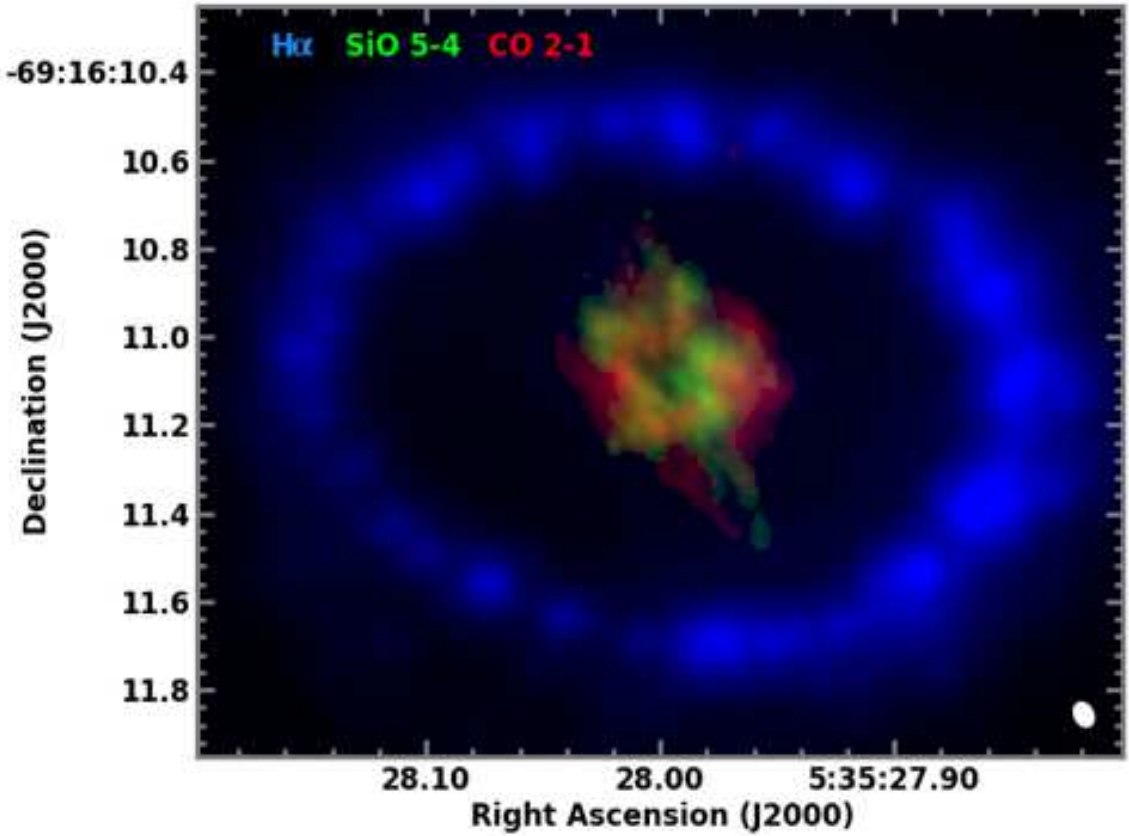}
\caption{Molecular emission and H$\alpha$ emission from SN 1987A. The more compact emission in the center of the image corresponds to the peak intensity maps of CO $2-1$ (red) and SiO $5-4$ (green) observed with ALMA. The surrounding H$\alpha$ emission (blue) observed with HST shows the location of the circumstellar equatorial ring \citep{Larsson16}. \label{fig:fig1}} 
\end{figure*}

\section{Observations and data reduction} \label{sec:observations}

ALMA observations of SN 1987A at 1.3 mm (Band 6, $211-275$ GHz) were performed in two different epochs: cycle 2 low angular resolution images were made in September 2, 2014 (A001/X10e/X140), and cycle 3 high angular resolution images were obtained from November 1 to 15, 2015 (A001/X1ee/X620). Combining data from both cycles improves the image quality. Both data sets were calibrated against QSOs (Quasi-stellar objects). 

For the cycle 2 data set J0519--4546 (05:19:49.72, -45:46:43.85; 0.75 Jy at 234 GHz) was the absolute flux calibrator and J0635--7516 (06:35:46.51, -75:16:16.82; 0.68 Jy at 234 GHz) was the phase calibrator. The cycle 3 high-resolution data consisted of two separate observations for each spectral setup. The $^{12}$C$^{16}$O $2-1$ (230.538 GHz) and $^{28}$Si$^{16}$O $5-4$ (217.105 GHz) images include data using J0519--4546 (0.75 Jy at 224 GHz) as the absolute flux calibrator and J0601--7036 (06:01:11.25, -70:36:08.79; 0.70 Jy at 224 GHz) as the phase calibrator. The SiO $6-5$ (260.518 GHz) image includes data using either J0519--4546 (0.63 Jy at 253 GHz) or J0334--4008 (03:34:13.65, -40:08:25.10; 0.44 Jy at 253 GHz) as the absolute flux calibrator and J0601--7036 (0.58 Jy at 253 GHz) as the phase calibrator.

Combining the cycle 2 and 3 data sets results in baselines between 34 m (25 k$\lambda$) and 16,196 m (12,600 k$\lambda$). We used the Common Astronomy Software Application (casa.nrao.edu) to process the interferometric data into a cube with 50 mas angular resolution and 100 km s$^{-1}$ spectral bins (for spherical expanding ejecta at the time of these observations, 100 km s$^{-1}$ corresponds to a path length of $\sim9\times10^{15}$ cm or 12 mas). For imaging and deconvolution we used the task \texttt{tclean} with different \texttt{multiscale} scales depending on the resulting beam sizes. For CO $2-1$ and SiO $5-4$ images, we adopted $scales=[0,7,21]$, and for the SiO $6-5$ image we adopted $scales=[0,5,15]$. In both cases, we used a 6 mas pixel size.
 
To determine the 3D distribution, we converted velocity into angular size on the sky by assuming the distance to the LMC to be 50 kpc. We adopted a spherical free expansion, and for the age of the supernova we took the time elapsed since the supernova explosion and the date of the observations ($\sim28.7$ years). The assumption of free expansion can be tested in the spatially summed image. Using the full width at zero intensity (FWZI) $\sim4000$ km s$^{-1}$, and a corresponding angular size of the velocity-integrated image of $\sim480$ mas, we obtain a time of $28.5$ years since the explosion. There is no evidence of deceleration of the supernova ejecta, and therefore the free expansion approximation is valid. 

\begin{figure*}[ht!]
\plotone{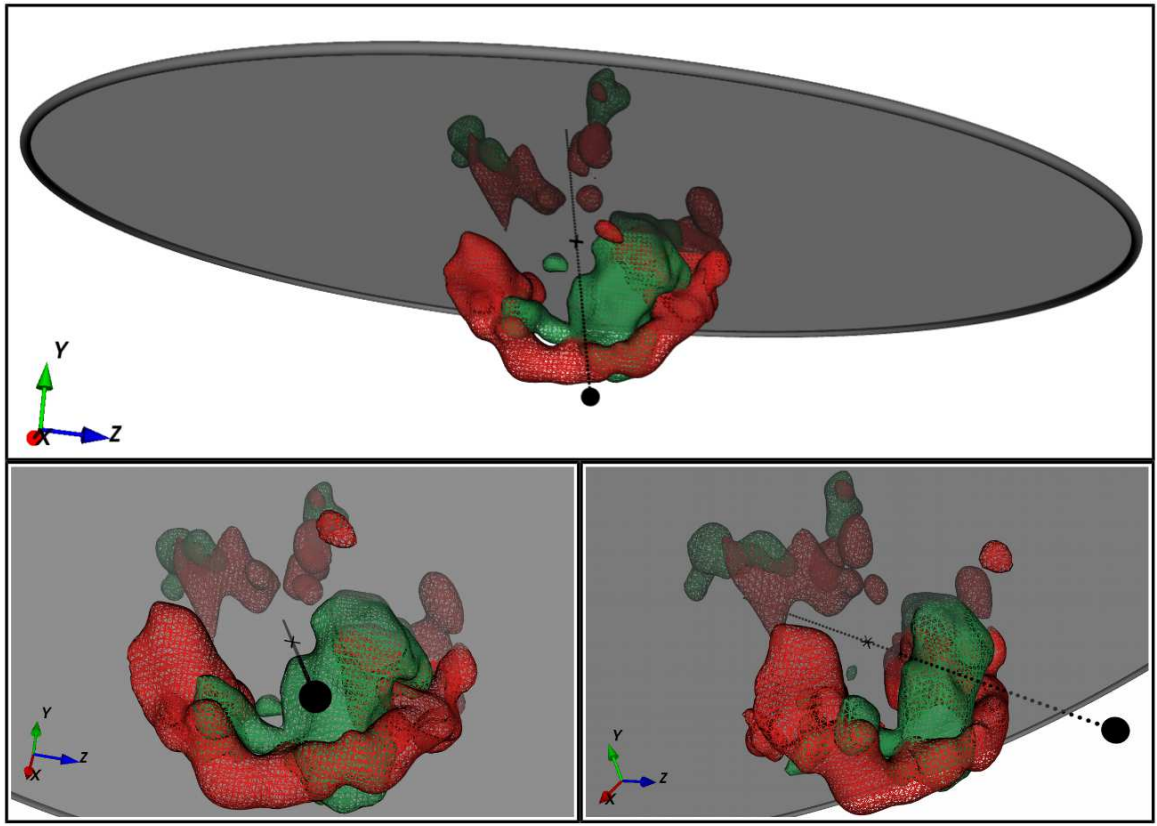}
\caption{3D view of cold molecular emission in SN 1987A. The CO $2-1$ (red) and SiO $5-4$ (green) emission is shown from selected view angles. The central region is devoid of significant line emission. The emission contours are at the $60\%$ level of the peak of emission for both molecules. The black dotted line and black filled sphere indicate the line of sight and the position of the observer, respectively. The gray ring shows the location of the reverse shock at the inner edge of the equatorial ring (XZ plane). The black cross marks the geometric center.\\ 
(A video of the 3D structure is available in the online journal) \label{fig:fig2}}
\end{figure*}

\section{Observational Results} \label{sec:results}

In Figure \ref{fig:fig1} we present the CO $2-1$ and SiO $5-4$ emission framed by the circumstellar equatorial ring. Phenomenologically the image provides strong clues to the structure of the inner ejecta of the supernova. The CO and SiO distributions are spatially distinct: both have a toroidal or shell-like distribution around the center, but most of the CO emission presents a maximum extension larger than SiO emission ($\pm1700$ km s$^{-1}$ vs. $\pm1300$ km s$^{-1}$; or $\pm1.53\times10^{17}$ cm vs. $\pm1.17\times10^{17}$ cm, approximately).

\begin{figure*}[ht!]
\plotone{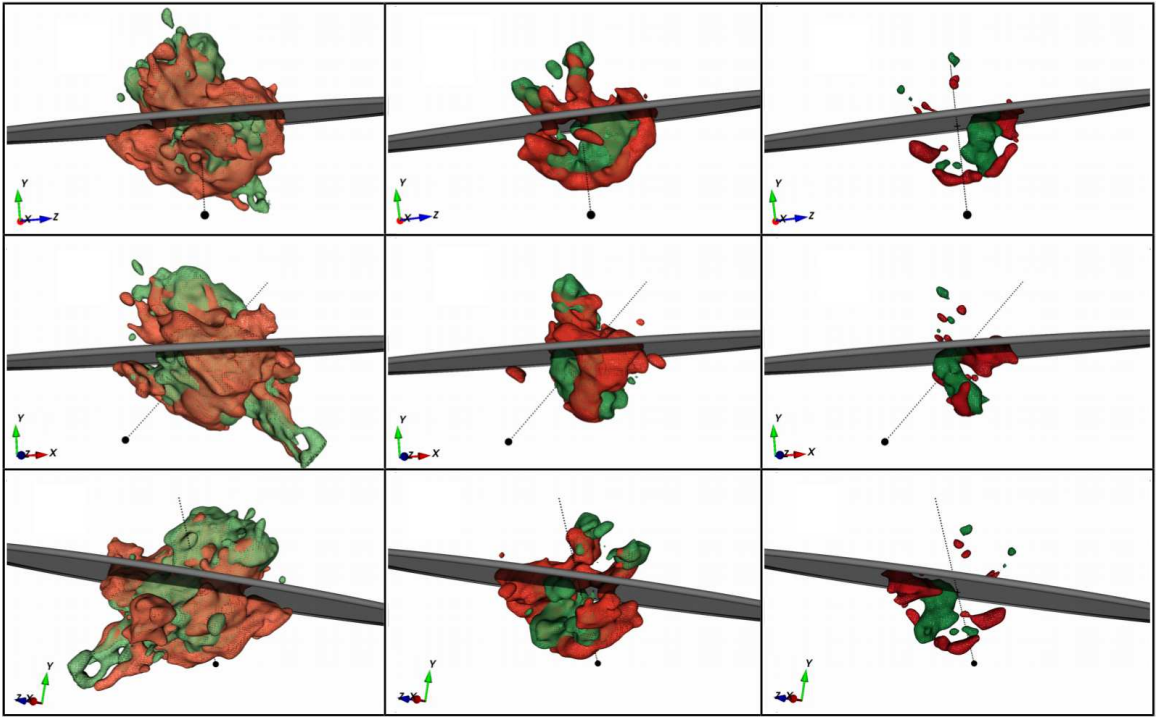}
\caption{3D distribution of emission of CO $2-1$ (red) and SiO $5-4$ (green) for selected view angles. From left to right the plots show the emission at $30\%$, $50\%$, and $70\%$ of the peak of emission of each line. \label{fig:fig3}} 
\end{figure*}

The 3D distribution of CO $2-1$ and SiO $5-4$ is shown in Figures \ref{fig:fig2} and \ref{fig:fig3}. Detailed examination reveals that the CO emission forms a torus-like shape perpendicular to the equatorial ring, clear evidence of asymmetry in the explosion. By contrast, SiO is clumpier and distributed in a broken shell rather than a torus. Within a region $\leq1500$ km s$^{-1}$ ($1.35\times10^{17}$ cm) from the center of the explosion, we find that 25\% of the clumps have peaks brighter than $1.9\times$ and $1.6\times$ the mean CO and SiO intensity, respectively. Translating the peak brightness of clumps into peak density of clumps requires modeling the non-LTE excitation of each clump, which we cannot do with only one emission line. However, if we use the average kinetic temperature and collider density derived for the entire nebula \citep{Matsuura17}, then a peak brightness $3\times$ brighter than average would translate into an SiO density $5\times$ higher than average, for the typical clump size.

The brightest SiO emission is from a single blob located off-center below the equatorial plane (i.e., the side located closer to Earth). In both species, the emission joins into a more continuous torus or shell at about $50\%$ of the peak of emission, and the central deficit of emission becomes fully enclosed at about $30\%$ of the peak of emission (see Figure \ref{fig:fig3}). 

The observed low emission at the central region agrees with predictions from numerical models according to which different mechanisms \citep[e.g., $^{56}$Ni-bubble effect, second outward shock,][]{Ertl16} would accelerate the innermost material outwards. The zones with the lowest emission ($<20\%$ of peak of emission) have similar sizes as the brightest clumps ($>70\%$ of peak of emission) and a similar filling fraction of $\sim25\%$. The parts of the ejecta with the least SiO emission are preferentially found around the center ($r\leq0\farcs1$ or $r\leq 7.5\times10^{16}$ cm), whereas holes of emission in CO span a range of radii similar to the range spanned by the brightest CO clumps. We have examined the distribution of these zones of minimal emission and find that both CO and SiO holes are roughly perpendicular to the CO torus. One possibility could be that these regions are filled with material of different composition, like the heavy-element-dominated $^{56}$Ni fingers predicted by some models.

The structure of the molecular emission is not aligned with the emission of 1.644 $\mu$m [Si I]+[Fe II], which is concentrated in two asymmetric lobes fairly close to the plane of the ring and is brightest at larger radii than both CO and SiO \citep{Larsson16}. Differences could be due to chemistry, molecular dissociation by positrons, different excitation mechanisms (thermal versus non-thermal), or dust obscuration. Figures \ref{fig:fig2} and \ref{fig:fig3} also reveal the first direct evidence of non-spherical instabilities, as SiO extends to greater radial distances (velocities) than CO in some directions. Considering emission brighter than $30\%$ of the peak of emission, SiO has a greater extent than CO in $25\%$ of radial directions, and in a few directions, extends up to 500 km s$^{-1}$ greater radial velocities.

\section{Comparison with hydrodynamical models} \label{sec:comparisons}

We compare our data to 3D numerical models of neutrino-driven core-collapse supernovae obtained with the finite-volume Eulerian multifluid hydrodynamics code PROMETHEUS \citep{Utrobin15, Wongwathanarat15} that employs an approximate, gray neutrino transport. All models considered are single-star models. To follow the simulations from $11-15$ milliseconds after core-bounce to about 150 days we also include a description of the $^{56}$Ni heating. The additional heating increases the outward velocities of the innermost $^{12}$C and $^{28}$Si by about $20\%$, so while it is not likely the only cause of the observed central evacuation, this ``nickel heating'' effect does contribute. 

\begin{figure*}[ht!]
\plotone{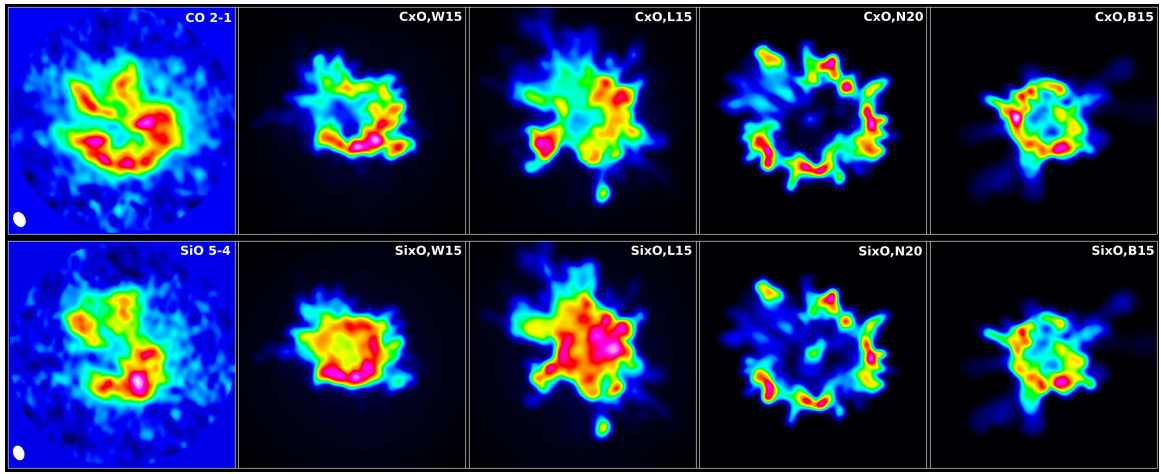}
\caption{Top row: a single slice $\sim4.5\times10^{15}$ cm thick through the center of the 3D distributions of observed CO $2-1$ and modeled $^{12}$C$\times^{16}$O for W15, L15, N20 and B15 at $75\%$ of peak CO emission or of peak $^{12}$C$\times^{16}$O density. Bottom row: same as top row, but for observed SiO $5-4$ and the same models for $^{28}$Si$\times^{16}$O at $65\%$ of peak SiO emission or of peak $^{28}$Si$\times^{16}$O density. The dynamics of the explosion of these models was not tailored to match the properties of SN 1987A. Thus, while clumpiness, radial extent, and length scales can be compared, an exact morphological match would be merely coincidental. The clump sizes and distribution of the models are visually similar from different viewing angles and do not depend on the particular orientation of the slice. In these plots, the CO and SiO data cubes have been rotated 43$^{\circ}$ so that the equatorial ring is horizontal.\label{fig:fig4}}
\end{figure*}

The progenitor mass for SN 1987A is in the range of $14-20$ $M_\odot$ \citep{Arnett89, McCray93, Smartt09}. We study four different non-rotating, non-magnetic progenitors: two 15 $M_\odot$ red supergiants (RSGs), model W15 \citep{Woosley95} and model L15 \citep{Limongi00}, and two blue supergiants (BSGs) tailored to represent SN 1987A progenitors: a 20 $M_\odot$ model N20 \citep{Shigeyama90} and a 15 $M_\odot$ model B15 \citep{Woosley88}. The BSG models have a shallower density profile at the H/He interface than the RSG models, and differ significantly in the C+O core mass (N$20\sim3.8$ $M_\odot$ vs. B$15\sim1.7$ $M_\odot$). 

One-dimensional models of molecule formation in supernovae have been published \citep{Lepp90,Liu96,Sarangi13,Sluder16} but there are no models that combine higher-dimensional hydrodynamic evolution of the ejecta with chemical processes. The emitted line radiation also depends on molecular excitation. This deep in the ejecta, it is reasonable to assume that excitation and heating are predominantly due to $^{44}$Ti decay. CO and SiO emission may therefore reflect the $^{44}$Ti abundance distribution in addition to the CO and SiO abundance distributions. We have not yet detected variations in the CO or SiO excitation temperature which would result from inhomogeneous excitation, but such variations may be detectable with future analysis of ALMA data. 

Nevertheless, in order to focus on the size scales and spatial distributions, we compare the CO and SiO emission distributions directly to the square root of the product of the $^{12}$C and $^{16}$O, and of $^{28}$Si and $^{16}$O modeled atomic density distributions. We also analyzed $^{12}$C and $^{28}$Si individually, with and without density ceilings at which CO and SiO become optically thick. If the yields are high (a large fraction of the available atoms remains in gas-phase molecules), then the molecular densities will be high enough to make the observed CO and SiO lines optically thick. 

We tested this in our structure comparisons by introducing an upper cutoff to the model volume density. For a rotational transition from level $J$ at energy $E_u$ to level $J-1$ at energy $E_l$, the Sobolev large velocity gradient optical depth is related to the molecular density $n$, rotational partition function $Q$ and excitation temperature $T_x$. For a diatomic rotator, like CO or SiO, the energy levels are $E_J = BJ(J + 1)$ where $B$ is the rotational constant and $g_J = 2J + 1$ is the statistical weight. As long as $T_x$ is not too low compared to $B$, the partition function can be approximated by $Q\sim T/B$, with $B/k_{B}=2.766$ K. Since the kinetic temperature is $T\sim50$ K \citep{Matsuura17}, the approximation is very reasonable. Assuming the same $T_x$ for all levels, 



\begin{equation}
\tau \approx \frac{\lambda^3 A_{ul} g_{u}t}{8\pi Q}\left(e^{-E_{l}/T_{x}} - e^{-E_{u}/T_{x}}\right)n
\end{equation}

For the CO $2-1$ transition, the Einstein coefficient $A_{21}=6.9\times10^{-7}$ s$^{-1}$. At $T\sim50$ K, and at the time of the observation, $\tau$\textsubscript{CO} $\approx 2.7\times10^{-3}$ $n$\textsubscript{CO}. Optical depth unity therefore corresponds to a CO density of $n$\textsubscript{CO}$\sim374$ cm$^{-3}$. This is within a factor of order unity of the maximum model densities, so does not affect our conclusions about relative structure size distributions. The same calculation for the SiO $5-4$ transition, with $A_{54}=5.2\times10^{-4}$ s$^{-1}$, yields $\tau$\textsubscript{SiO} $\approx 4.2$ $n$\textsubscript{SiO}. Thus, optical depth unity corresponds to a SiO density of $n$\textsubscript{SiO} $\sim 0.24$ cm$^{-3}$, a rather more severe threshold compared to model densities. From lower angular resolution ALMA observations, the total CO and SiO masses are a few $\times 10^{-2}$ $M_{\odot}$ and a few $\times 10^{-4}$ $M_{\odot}$, respectively, with uncertainties of a factor of several \citep{Matsuura17}. These masses account for $<20\%$ and $<0.1\%$ of the expected C and Si yields \citep{Woosley95}. Models suggest that formation of silicate dust is very efficient in this environment, leaving only a small fraction of $^{28}$Si currently in molecular SiO \citep{Sarangi13}. Thus, until more complete 3D models including dust formation exist, we compare the modeled $^{28}$Si$\times^{16}$O distribution without a cutoff with observed SiO.

\begin{figure*}
\gridline{\fig{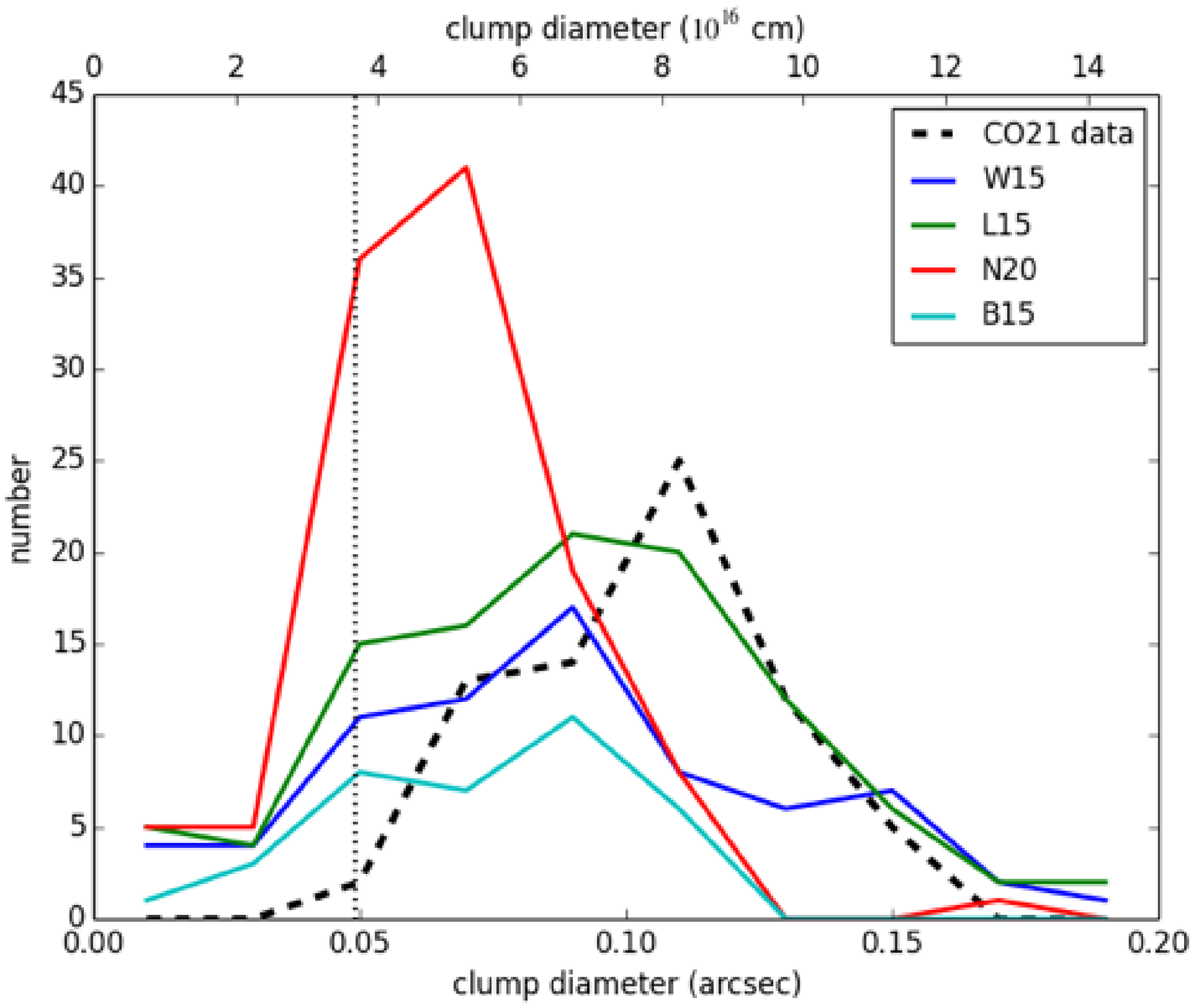}{0.4\textwidth}{}
          \fig{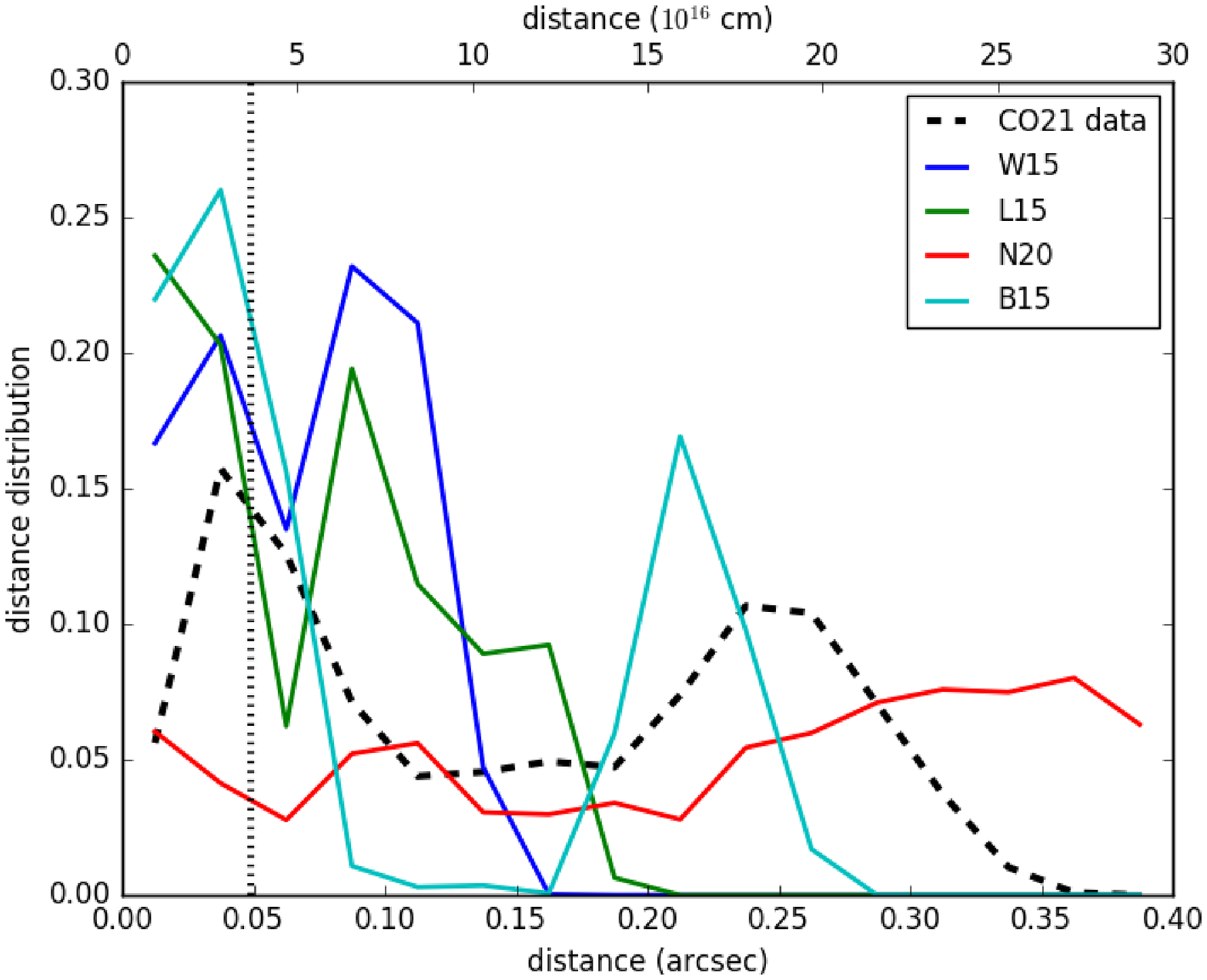}{0.4\textwidth}{}
          }
\gridline{\fig{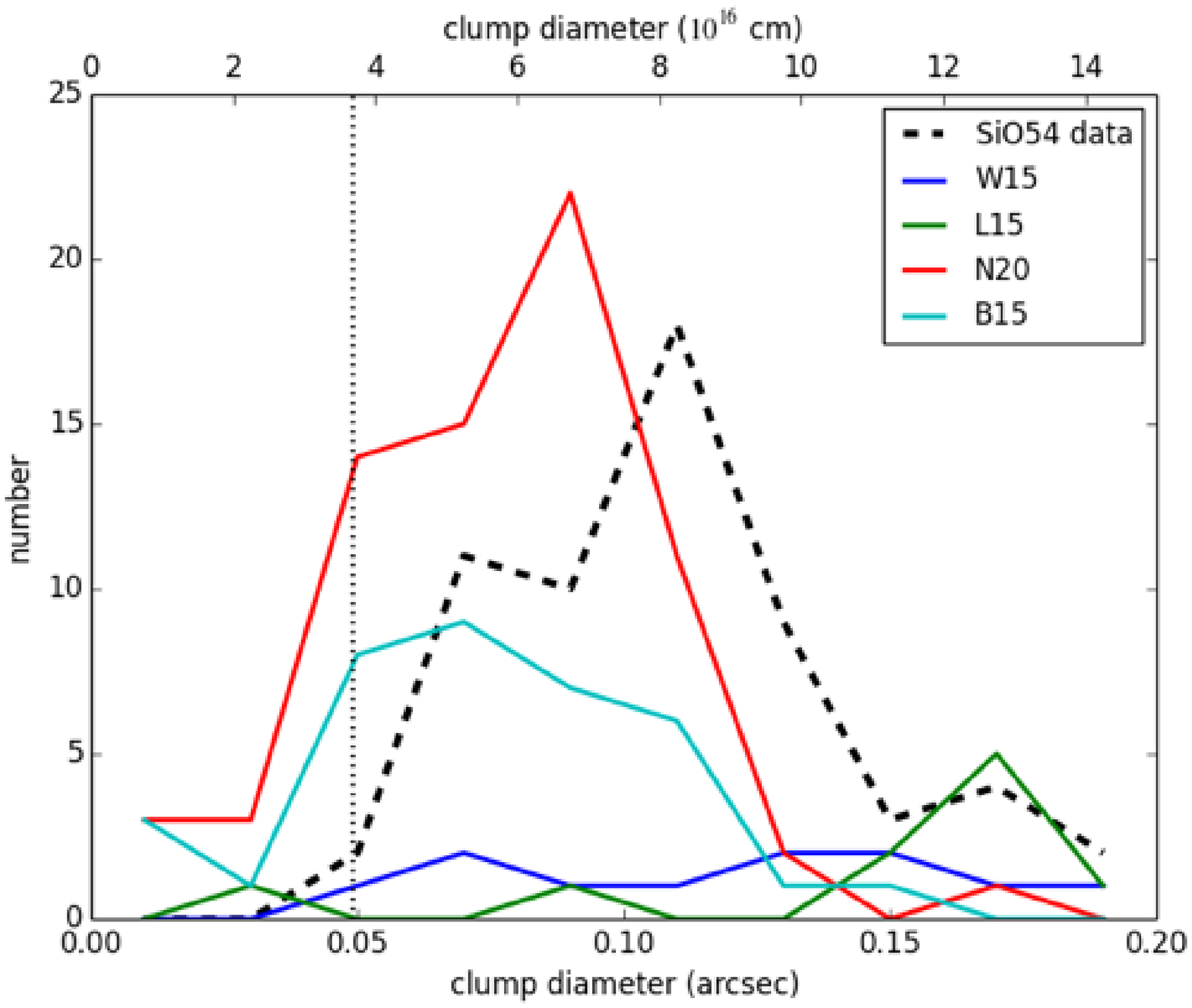}{0.4\textwidth}{}
          \fig{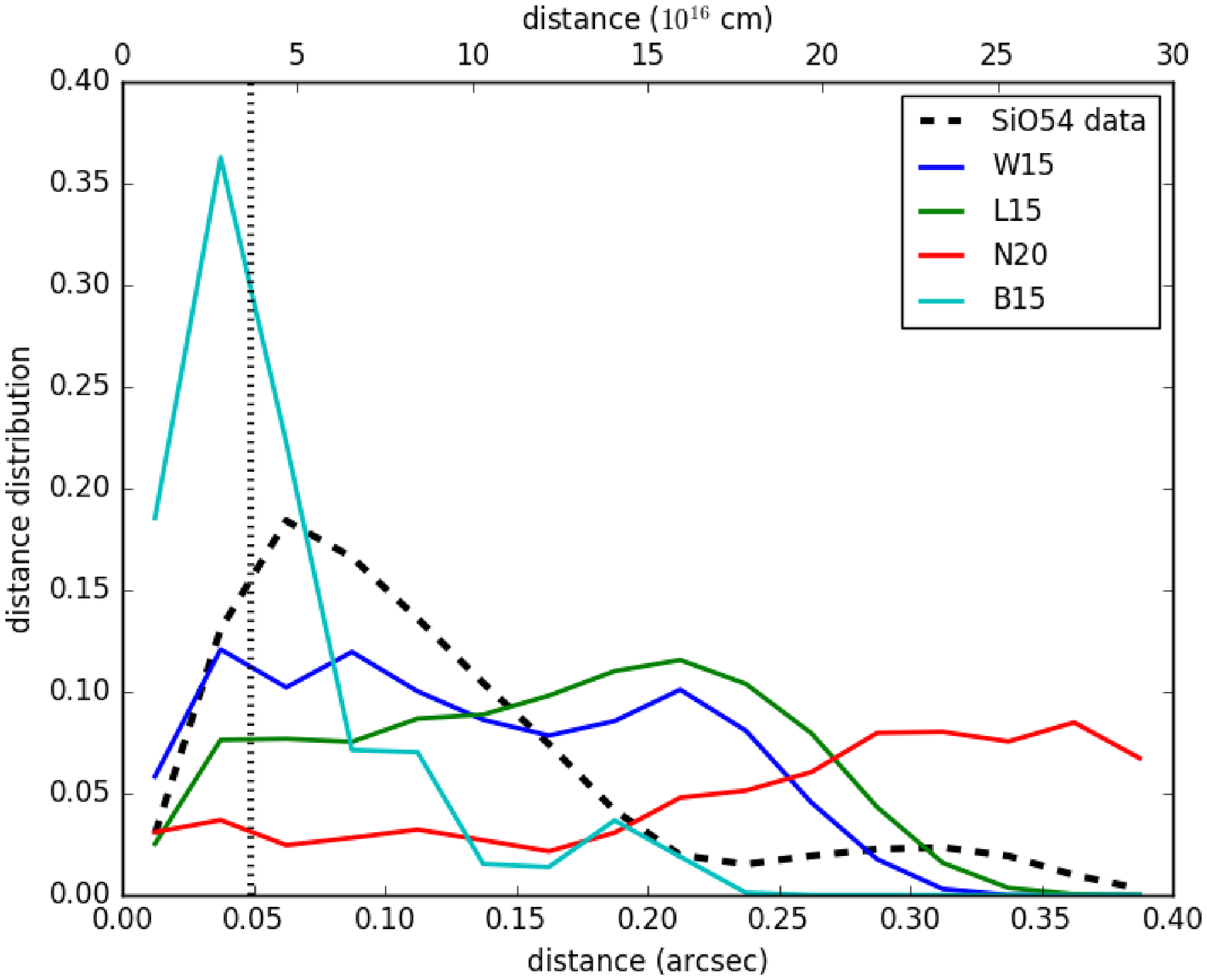}{0.4\textwidth}{}
          }
\gridline{\fig{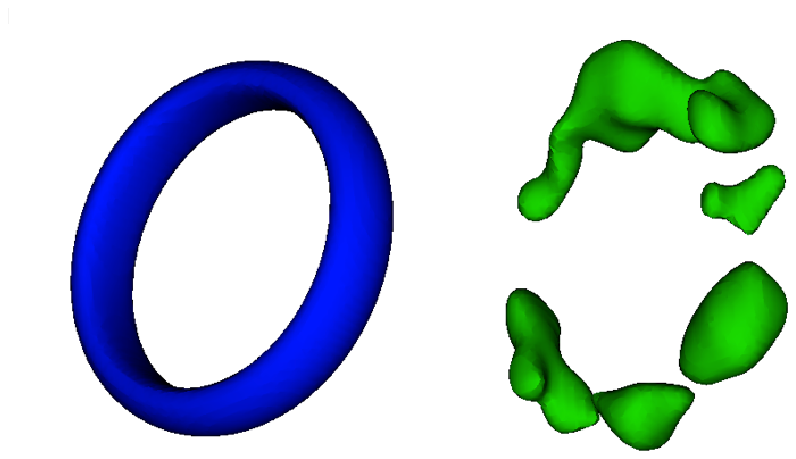}{0.4\textwidth}{}
          \fig{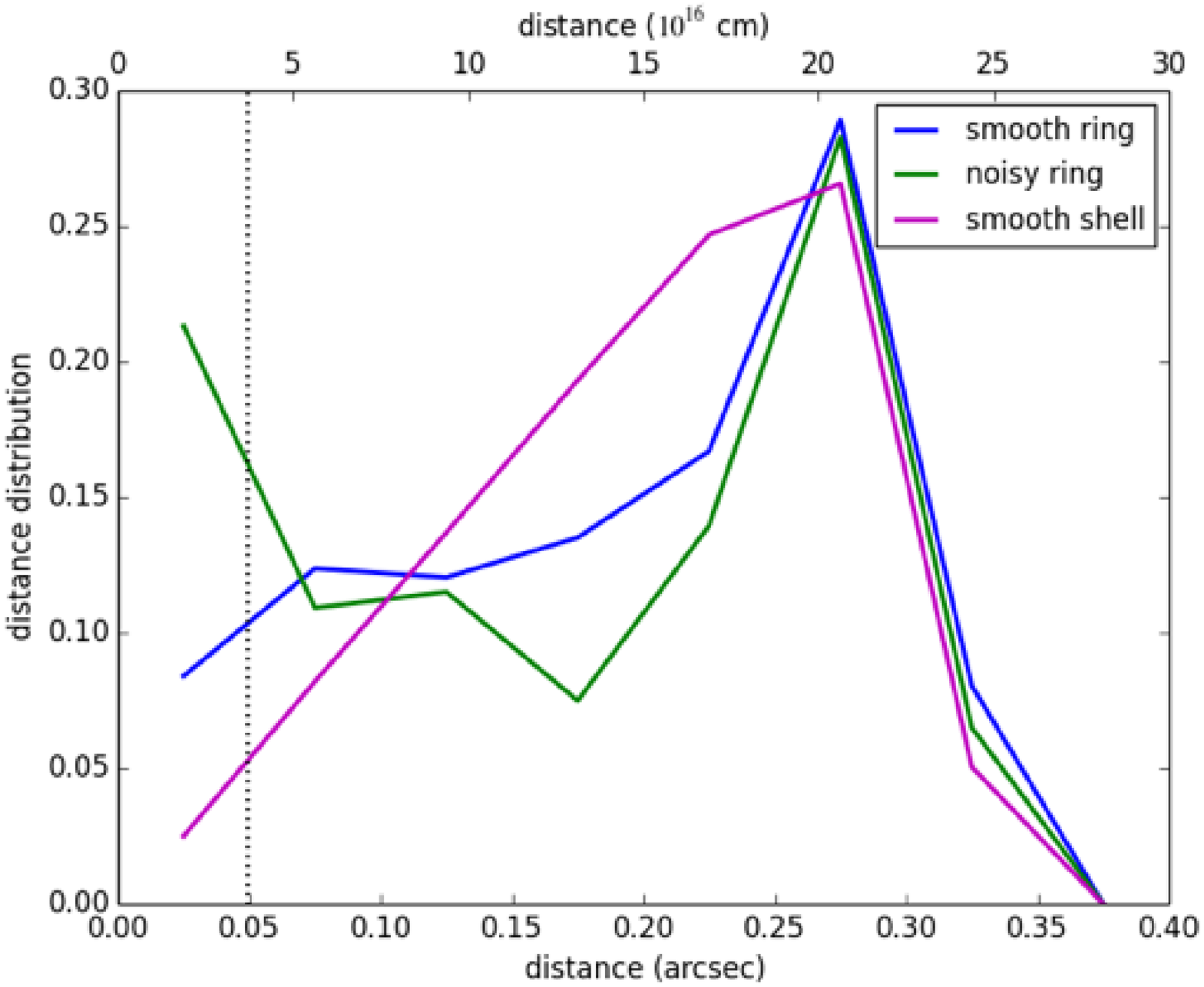}{0.4\textwidth}{}
          }
\caption{Top left: distribution of clump sizes, in CO emission and in the product of $^{12}$C$\times^{16}$O modeled density. Top right: clump separation between points with bright emission, or high modeled atomic density. All points in the 3D cubes above $75\%$ of peak CO emission, or of peak $^{12}$C$\times^{16}$O density for the models, are included in the calculation. The CO data show a clear peak at $\sim0\farcs25$ consistent with a toroidal structure of that diameter. Center left: distribution of clump sizes, in SiO emission and in the product of $^{28}$Si$\times^{16}$O modeled density. Center right: same as top right, but for SiO at $65\%$ of peak SiO emission, or of peak $^{28}$Si$\times^{16}$O density for the models. The SiO data show a minor peak at $\sim0\farcs30$ consistent with a broken shell structure of that diameter. Bottom left: two simple shapes, (a) a ring or torus, and (b) clumps arranged in a ring-like distribution similar to observed in CO. Bottom right: clump separation of (a) shows a strong peak at the diameter of the ring, while the clump separation of (b) maintains that peak but in addition shows a peak at small scales tracing typical clump sizes and separations. A closed shell (not shown) shows a peak at the diameter of the shell, but a broader range of intermediate distances. In all plots, the vertical dotted line represents the beam size.
\label{fig:fig5}}
\end{figure*}

The results obtained for the clump sizes and distributions are robust to the different approaches. The main difference is that $^{28}$Si$\times^{16}$O has a more pronounced central hole than $^{28}$Si alone, more similar to the SiO data. Due to the fact that $^{16}$O extends further out than $^{28}$Si, their intersection has more of a hole. Instead, the differences between $^{12}$C$\times^{16}$O and $^{12}$C alone are less significant, as expected since these nucleosynthetic products start out at more similar radii before the instabilities take place. Figure \ref{fig:fig4} shows the phenomenological comparison of the data and the four models homologously scaled from 150 days to 29 years, and convolved to the observational resolution. 

Different quantitative metrics are sensitive to different aspects of a 3D hierarchical structure. Segmenting the cube into clumps using \textit{cprops} \citep{Rosolowsky06} and \textit{clumpfind} \citep{Williams94} algorithms reveals the clump size distribution. On the other hand, the radially averaged 3D Fourier transform, and the distribution of distances between pairs of points above a given intensity threshold (thresholds between $50\%$ and $90\%$ of peak of emission were considered), reveal the range of clump separations and any large-scale structure such as the toroidal arrangement of CO. Figure \ref{fig:fig5} shows the clump size and clump separation for data and models of CO and SiO. In order to help illustrate the metric, simple geometries are also shown in Figure \ref{fig:fig5}. 

The results that are most robust to different thresholds and segmentation parameters are the following: none of the models fit both emission lines in all structural metrics, but for CO, the RSG models W15 and L15 have $^{12}$C$\times^{16}$O clump sizes comparable to the data. Instead, for SiO, the RSG models have a large range of $^{28}$Si$\times^{16}$O clump sizes, on average, $\sim35$\% larger than the typical SiO clump sizes. For both lines, the clump sizes of BSG models N20 and B15 tend to be too small compared to the observations.

The clump separation of CO shows a clear peak at $\sim0\farcs25$ consistent with a toroidal structure of that diameter, while the distribution of SiO shows a minor peak at $\sim0\farcs30$. For both CO and SiO, the RSG models do not have such a clear peak, but instead they have a broader range of intermediate scales, reflecting the intermediate-sized structures visible in Fig. \ref{fig:fig4}. For SiO, both W15 and L15 models have a large-scale shell or torus comparable in extent to the data. The BSG models are more fragmented into small structures than the observations. Model N20 has small clumps with a wide range of clump-to-clump distances, and a larger shell structure than the data. For CO, the BSG model B15 does not have a range of intermediate distances but shows a strong peak at $\sim0\farcs22$, more similar to the CO data. 

\section{Conclusions} \label{sec:conclusions}

We have used the extraordinary capabilities of ALMA to create unique 3D maps of CO and SiO molecular emission from the inner ejecta of SN 1987A with unprecedented spatial resolution. Our observations show the clumpy mixed structure predicted by models \citep[e.g.,][]{Hammer2010} but that were not previously well imaged, as well as a clear sign of asymmetry in the supernova explosion. We find that the molecular emission forms a torus-/shell-like shape perpendicular to the equatorial ring, and in some directions, the SiO extends further out than CO from the center of the remnant, a proof that non-spherical instabilities have taken place at the time of the explosion. 

From the comparison of these data to hydrodynamical models we conclude that none of the models correctly reproduce neither the radial extent of both lines nor the range of clump sizes. The first discrepancy might be solved by adjusting the explosion energy of the model, but the second is directly related to the progenitor structure and the strongest asymmetries produced during shock revival, or to as of yet unmodeled chemical or excitation effects. However, considering that the models are not particularly tuned to reproduce SN 1987A, the agreement in some features as the overall shape of CO and SiO emission is surprisingly good.

The models compared here are non-rotating, and thus we cannot say how different the signatures of SASI would be. Comparing these data with a larger range of models, also including SASI mechanisms, will constrain explosion physics more precisely than has been possible heretofore using unresolved images and spectra, as different model parameters affect the size scales of clumpy structure in different ways: the explosion energy changes the overall extent, instability physics and the duration of the unstable phase affects the small-scale clump sizes. Progenitor structure and astrochemistry (fractional molecular yield compared to elemental abundance) also change the range of clump sizes and extent of their envelopes. 

\acknowledgments
This Letter makes use of the following ALMA data: ADS/JAO.ALMA$\#$ 2013.1.00280.S and ADS/JAO.ALMA$\#$ 2015.1.00631.S. ALMA is a partnership of ESO (representing its member states), NSF (USA) and NINS (Japan), together with NRC (Canada), NSC and ASIAA (Taiwan), and KASI (Republic of Korea), in cooperation with the Republic of Chile. The Joint ALMA Observatory is operated by ESO, AUI/NRAO and NAOJ. The National Radio Astronomy Observatory is a facility of the National Science Foundation operated under cooperative agreement by Associated Universities, Inc. F.J.A. and J.M.M. were partially supported by the Spanish MINECO projects AYA2012-38491-C02-01 and AYA2015-63939-C2-2-P and the Generalitat Valenciana projects PROMETEO/2009/104 and PROMETEOII/2014/057. At Garching, funding by the European Research Council through grant ERC-AdG No. 341157-COCO2CASA is acknowledged. P.C. acknowledges support from the European Research Council (ERC) in the form of Consolidator Grant CosmicDust (ERC-2014-CoG-647939, PI H.L. Gomez). J.L. acknowledges support from the Knut and Alice Wallenberg foundation. M.M. is supported by an STFC Ernest Rutherford fellowship (ST/L003597/1). S.W. acknowledges support from NASA (NNX14AH34G).

\listofchanges

\begin{thebibliography}{}


\bibitem[Arnett et al.(1989)]{Arnett89} Arnett, W. D., Bahcall, J. N., Kirshner, R. P., \& Woosley, S. E. 1989, ARA\&A, 27, 629
\bibitem[Basko(1994)]{Basko94} Basko, M. 1994, ApJ, 425, 264
\bibitem[Benz \& Thielemann(1990)]{Benz90} Benz, W., $\&$ Thielemann, F.-K. 1990, ApJ, 348, L17 
\bibitem[Blondin et al.(2003)]{Blondin03} Blondin, J. M., Mezzacappa, A., $\&$ DeMarino, C. 2003, ApJ, 584, 971 
\bibitem[Burrows et al.(1995)]{Burrows95} Burrows, A., Hayes, J., $\&$ Fryxell, B. A. 1995, ApJ, 450, 830
\bibitem[Ebisuzaki et al.(1989)]{Ebisuzaki89} Ebisuzaki, T., Shigeyama, T., $\&$ Nomoto, K. 1989, ApJ, 344, L65
\bibitem[Ertl et al.(2016)]{Ertl16} Ertl, T., Ugliano, M., Janka, H.-Th., Marek, A., Arcones, A. 2016, ApJ, 821, 69 
\bibitem[Fransson et al.(2013)]{Fransson13} Fransson, C., Larsson, J., Spyromilio, J., et al. 2013, ApJ, 768, 88
\bibitem[Haas et al.(1990)]{Haas90} Haas, M. R., Colgan, S. W. J., Erickson, E. F., et al. 1990, ApJ, 360, 257 
\bibitem[Hammer et al.(2010)]{Hammer2010} Hammer, N. J., Janka, H.-T., \& Müller, E. 2010, ApJ, 714, 1371
\bibitem[Herant \& Benz(1991)]{Herant91} Herant, M., \& Benz, W. 1991, ApJ, 370, L81
\bibitem[Herant \& Benz(1992)]{Herant92} Herant, M., \& Benz, W. 1992, ApJ, 387, 294
\bibitem[Jakobsen et al.(1991)]{Jakobsen91} Jakobsen, P., Albrecht, R., Barbieri, C., et al. 1991, ApJ, 369, L63 
\bibitem[Kamenetzky et al.(2013)]{Kamenetzky13} Kamenetzky, J., McCray, R., Indebetouw, R., et al. 2013, ApJL, 773, L34 
\bibitem[Larsson et al.(2011)]{Larsson11} Larsson, J., Fransson, C., \"Ostlin, G., et al. 2011, Nature, 474, 484
\bibitem[Larsson et al.(2016)]{Larsson16} Larsson, J., Fransson, C., Spyromilio, J., et al. 2016, ApJ, 833, 147 
\bibitem[Lawrence et al.(2000)]{Lawrence00} Lawrence, S. S., Sugerman, B. E., Bouchet, P., et al. 2000, ApJL, 537, L123
\bibitem[Lepp et al.(1990)]{Lepp90} Lepp, S., Dalgarno, A., $\&$ McCray, R. 1990, ApJ, 358, 262
\bibitem[Limongi et al.(2000)]{Limongi00} Limongi, M., Straniero, O., $\&$ Chieffi, A. 2000, ApJS, 129, 625
\bibitem[Liu \& Dalgarno(1996)]{Liu96} Liu, W., \& Dalgarno, A. 1996. ApJ. 471, 480
\bibitem[Matsuura et al.(2015)]{Matsuura15} Matsuura, M., Dwek, E., Barlow, M. J., et al. 2015, ApJ, 800, 50
\bibitem[Matsuura et al.(2017)]{Matsuura17} Matsuura, M., Indebetouw, R., Woosley, S., et al. 2017, arXiv:1704.02324 
\bibitem[McCray(1993)]{McCray93} McCray, R. 1993, ARA$\&$A, 31, 175 
\bibitem[McCray \& Fransson(2016)]{McCray16} McCray, R., \& Fransson, C. 2016 ARA\&A, 54, 19
\bibitem[Rank et al.(1988)]{Rank88} Rank, D. M., Pinto, P. A., Woosley, S. E., Bregman, J. D., \& Witteborn, F. C. 1988, Nature, 331, 505
\bibitem[Roche et al.(1991)]{Roche91} Roche, P. F., Aitken, D. K., \& Smith, C. H. 1991, MNRAS, 252, 39P
\bibitem[Rosolowsky \& Leroy(2006)]{Rosolowsky06} Rosolowsky, E., \& Leroy, A. 2006, PASP, 118, 590
\bibitem[Sarangi \& Cherchneff(2013)]{Sarangi13} Sarangi, A., \& Cherchneff, I. 2013, ApJ, 776, 107 
\bibitem[Shigeyama $\&$ Nomoto(1990)]{Shigeyama90} Shigeyama, T., $\&$ Nomoto, K. 1990, ApJ, 360, 242
\bibitem[Sluder et al.(2016)]{Sluder16} Sluder, A., Milosavljevic, M., \& Montgomery, M. H. 2016, arXiv:1612.09013
\bibitem[Smartt(2009)]{Smartt09} Smartt, S. J. 2009, ARA$\&$A, 47, 63 
\bibitem[Sonneborn et al.(1998)]{Sonneborn98} Sonneborn, G., Pun, C. S. J., Kimble, R. A., et al. 1998, ApJL, 492, L139
\bibitem[Spyromilio et al.(1988)]{Spyromilio88} Spyromilio, J., Meikle, W. P. S., Learner, R. C. M., \& Allen, D. A. 1988, Nature, 334, 327
\bibitem[Spyromilio et al.(1990)]{Spyromilio90} Spyromilio, J., Meikle, W. P. S., \& Allen, D. A. 1990, MNRAS, 242, 669
\bibitem[Tziamtzis et al.(2011)]{Tziamtzis11} Tziamtzis, A., Lundqvist, P., Gröningsson, P., \& Nasoudi-Shoar, S. 2011, A\&A, 527, A35
\bibitem[Utrobin et al.(2015)]{Utrobin15} Utrobin, V. P., Wongwathanarat, A., Janka, H.-Th., \& M\"uller, E. 2015 A$\&$A, 581, A40
\bibitem[Williams et al.(1994)]{Williams94} Williams, J. P., de Geus, E. J., \& Blitz, L. 1994, ApJ, 428, 693
\bibitem[Wongwathanarat et al.(2015)]{Wongwathanarat15} Wongwathanarat, A., M\"uller, E., $\&$ Janka, H.-T. 2015, A$\&$A, 577, A48 
\bibitem[Woosley et al.(1988)]{Woosley88} Woosley, S. E., Pinto, P. A., $\&$ Ensman, L. 1988, ApJ, 324, 466
\bibitem[Woosley $\&$ Weaver(1995)]{Woosley95} Woosley, S. E., $\&$ Weaver, T. A. 1995, ApJS, 101, 181
\end{thebibliography}
\end{document}